\documentclass[epj]{svjour}

\usepackage{psfig}
\usepackage{graphicx}
\usepackage{dcolumn}
\usepackage{bm}

\usepackage{amssymb}

\begin{document}

\title{Spin dynamics and coherent tunnelling in the molecular magnetic rings 
Fe$_6$ and Fe$_8$}

\author{A. Honecker\inst{1} \and F. Meier\inst{2} \and Daniel Loss\inst{2} 
\and B. Normand\inst{3}\thanks{\emph{Present address:} D\'epartement de 
Physique, Universit\'e de Fribourg, CH-1700 Fribourg, Switzerland}}

\institute{Institut f\"ur Theoretische Physik, TU Braunschweig, 
Mendelssohnstr. 3, D-38106 Braunschweig, Germany \and Departement f\"ur 
Physik und Astronomie, Universit\"at Basel, CH-4056 Basel, Switzerland 
\and Theoretische Physik III, Elektronische Korrelationen und 
Magnetismus, \\ Institut f\"ur Physik, Universit\"at Augsburg, D-86135 
Augsburg, Germany }

\date{Received: Feb.~22nd, 2002}

\abstract{
We present detailed calculations of low-energy spin dynamics in the ``ferric 
wheel'' systems Na:Fe$_6$ and Cs:Fe$_8$ in a magnetic field. We compute by 
exact diagonalisation the low-energy spectra and matrix elements for 
total-spin and N\'eel-vector components, and thus the time-dependent 
correlation functions of these operators. Comparison of our results with 
the semiclassical theory of coherent quantum tunnelling of the N\'eel 
vector demonstrates the validity of a two-state description for the 
low-energy dynamics of ferric wheels. We discuss the implications of 
our results for mesoscopic quantum coherent phenomena, and for the 
experimental techniques to observe them, in molecular magnetic rings.
}


\maketitle

\section{Introduction}

\label{secI}

Quantum coherent tunnelling of the magnetic moment in nanoscopic magnets 
has recently become the focus of strong experimental and theoretical 
activity \cite{rgb}. The ferric wheel systems Fe$_N$ (Fig.~1) present a 
particularly promising subgroup in which crystals have now been prepared 
of compounds with $N = 6, 8, 10, 12$ and 18 magnetic Fe(III) ions in 
ring geometry \cite{rccffggs,rcajag,rsbuh,rgcps,rtdpfgl,rccfg,rwfpccal}. 
These molecules have antiferromagnetic (AF) coupling between spins $s = 5/2$ 
on each iron site, show a ground state with vanishing total spin $S = 0$ at 
zero field, and because of an effective uniaxial magnetic anisotropy admit 
the possibility of mesoscopic quantum phenomena in the form of coherent 
tunnelling of the N\'eel vector \cite{rbc,rkz,rldg,rcl}.

The best characterised molecular rings are Fe$_{10}$ \cite{rgcps,rtdpfgl}, 
various realisations of Fe$_6$ \cite{rccffggs,rcajag}, which differ in 
ligand group and central alkali metal ion, and Cs:Fe$_8$ \cite{rsbuh}. 
These materials have been studied by a variety of experimental 
techniques, including magnetic susceptibility and torque 
magnetometry \cite{rccffggs,rcja,rwskmbsaga,rwkssmbshb}, specific 
heat \cite{ralcc}, electron spin resonance (ESR) \cite{rpdkwg}, inelastic 
neutron scattering (INS) \cite{rwskmbsaga} and spin relaxation in nuclear
magnetic resonance (NMR) \cite{rjjlbhcg}. All of these studies serve 
essentially to characterise the zero-field spectrum and the dependence 
of the energetic separation of the lowest two levels on the applied field, 
both of which may be encapsulated within a phenomenological Hamiltonian 
written in terms of only the total spin \cite{rtdpfgl,rcja}.

\begin{figure}[t!]
\centerline{\psfig{figure=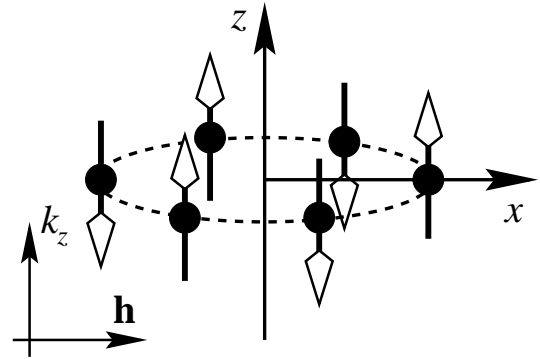,height=4.8cm,angle=0}}
\bigskip
\caption{Schematic representation of an Fe$_6$ ring with spins aligned 
along the easy axis (${\hat z}$), whose orientation is normal to the 
ring plane. The field ${\bf h} = g \mu_B {\bf B}$ is applied in 
the plane of the ring. }
\end{figure}

In contrast to the situation in the ferromagnetic (FM) molecular clusters 
Mn$_{12}$ and Fe$_8$ \cite{rwsll}, the notion of spin quantum tunnelling in 
the AF ferric wheels has to date received little experimental attention. On 
the theoretical side, a semiclassical description of the low-energy dynamics 
provides the clear prediction \cite{rcl} of coherent tunnelling of the 
staggered moment. Analysis of magnetisation and torque measurements in this 
framework \cite{rnwzl} allows one to extract the AF superexchange $J$ and 
an effective uniaxial anisotropy $k_z$ for the ring molecules. $k_z$ 
determines the height of the tunnel barrier between the degenerate energy 
minima, and thus the extent to which quantum transitions may influence the 
system response at the lowest temperatures. 

Although technical difficulties certainly arise in coupling directly to 
the staggered moment, the ferric wheels appear to offer significant 
advantages over FM molecules for the experimental observation of coherent 
tunnelling. In all such molecular crystals there are many possible sources 
of decoherence \cite{rnwzl} which act to destroy the coherent nature of the 
predicted tunnelling processes. These we discuss in further detail in Sec.~IV. 
Here we note only that any measurement of quantum coherence effects would 
require at minimum that the decoherence time $\Gamma^{-1}$ be significantly 
longer than the tunnelling time $\Delta^{-1}$. From the semiclassical theory, 
summarised in Sec.~\ref{secII}, $\Delta$ is the level separation in the
low-energy 
(two-state) manifold, and corresponds to the tunnel splitting. In the FM 
systems $\Gamma \gg \Delta$, and mesoscopic spin quantum tunnelling is said 
to be incoherent. While the reasonable assumption that decoherence rates, 
$\Gamma$, are not vastly different in the FM and AF systems remains to be 
proven, the tunnelling frequency, $\Delta$, is some 6-7 orders of magnitude 
larger in the AF ferric wheels, making these very good candidates for the 
observation of coherent tunnelling. We note here that none of the above 
types of experiment offers a means to extract a decoherence rate or to 
distinguish between coherent and incoherent tunnelling.

The decoherence rate can be determined only from dynamical 
quantities \cite{rml1}, whose spectral linewidths provide an upper bound 
on $\Gamma$. In the strictest sense, experimental confirmation of the 
inequality $\Gamma \ll \Delta$ still does not establish the existence 
of coherent dynamics, an undertaking which would require a true
time-domain observation of an appropriate oscillating quantity. In this 
study we seek to establish that coherent oscillations, including coherent 
tunnelling processes, are indeed present in the dynamical properties of 
ferric wheel systems without decoherence. That this is not a trivial 
statement is clear both phenomenologically from the apparent success 
of a model requiring only the total spin \cite{rtdpfgl,rcja}, and 
microscopically from the huge number of states [$(2 s + 1)^N$], 
energy-level splittings and possible matrix elements involved in a 
complete description. Establishment of the presence and nature of 
quantum coherence in these mesoscopic molecular systems would provide 
both an existence proof for coherent spin tunnelling if the condition 
$\Gamma \ll \Delta$ is satisified, and valuable guidance for its 
experimental observation. 

Thus we present a detailed investigation of the dynamical properties of 
the ferric wheels Na:Fe$_6$ and Cs:Fe$_8$ by exact diagonalisation (ED). 
A necessary initial step is to identify those dynamical quantities from 
which information on the quantum dynamics of interest may be obtained. 
This task is straightforward for FM molecular clusters, but is less 
immediately evident in ferric wheels, where, with the exception of a recent 
examination of electronic and nuclear spin dynamics in a semiclassical 
framework \cite{rml1}, a full analysis of microscopic dynamical properties 
is still lacking for the realistic Hamiltonian. The spin dynamics of rings 
with AF Heisenberg interactions has attracted some recent interest (see 
Refs.~\cite{rwk,rw} and references therein). However, previous studies
have been restricted to systems without anisotropy, which are not expected 
to show mesoscopic quantum phenomena in the form of coherent N\'eel vector
tunneling. 

In contrast, we consider here the spin dynamics of a full, effective 
Hamiltonian for the ferric wheels Fe$_6$ and Fe$_8$. The calculation of 
dynamical quantities, which we obtain from total-spin and N\'eel-vector 
correlation functions, then provides essential new information adding to 
the understanding of the physical properties of ferric wheels. Such a 
study is required not only to establish the existence of quantum coherent 
oscillations in a microscopic model for a complex, mesoscopic system, but 
also to aid the extraction of decoherence rates from experimental dynamical 
quantities. On the theoretical level, dynamical studies are qualitatively 
more difficult than the calculation of ground-state properties \cite{rnwzl} 
because they require that excited states and all corresponding matrix 
elements be taken into account. We note that a very recent, related study 
\cite{rrrs} of the systems Fe$_6$ to Fe$_{12}$, while technically advanced, 
includes the anisotropy only at the level of a low-energy effective 
Hamiltonian and does not consider intrinsic dynamical correlations in the 
presence of a time-independent magnetic field.

The manuscript is organised as follows. In Sec.~\ref{secII} we present the 
model Hamiltonian and observables, and provide a brief overview of the 
technique by which they are analysed. In Sec.~\ref{secIII} we present our 
results for Na:Fe$_6$ and Cs:Fe$_8$, both in the time domain and by analysis 
of matrix elements, and illustrate their physical origin by comparison with 
semiclassical approaches. In Sec.~\ref{secIV} we discuss the microscopic 
understanding of quantum coherence, decoherence sources and the prospects 
for experimental observation of spin tunnelling in ferric wheels. 
Sec.~\ref{secV} contains a summary and conclusions. 

\section{Model and Method}

\label{secII}

We work within the minimal model Hamiltonian for AF rings with effective 
uniaxial anisotropy \cite{rcl,rnwzl,rml1},
\begin{equation} 
H = J \sum_{i = 1}^N {\bf s}_i {\bf \cdot s}_{i+1} - k_z \sum_{i = 1}^N 
s_{i,z}^2 + {\bf h \cdot} \sum_{i = 1}^N {\bf s}_i ,
\label{emmh}
\end{equation}
where $N$ = 6 or 8, ${\bf s}_1 = {\bf s}_{N+1}$ and ${\bf h} = g \mu_B 
{\bf B}$. $J$ is the superexchange interaction which favours an 
AF spin configuration on the ring, $k_z$ is the effective uniaxial 
anisotropy, of dipolar and single-ion origin, and the final term is the 
Zeeman coupling. In the following all energies and fields are scaled to 
$J$ and $\hbar$ is set to unity. We define the total-spin operator
\begin{equation}
{\bf S} = \sum_i^N {\bf s}_i 
\label{ets}
\end{equation}
and the N\'eel-vector operator
\begin{equation}
{\bf n} = \frac{1}{N s} \sum_i^N (-1)^i {\bf s}_i .
\label{env}
\end{equation}
For finite anisotropy the total spin is not a good quantum number, meaning 
that $k_z$ leads to mixing of different spin multiplets, as will be evident 
from anticrossings in the energy spectra as a function of field to be shown 
below. This is a principal reason why numerical calculations are required 
to make further progress in a quantitative analysis of Eq.~(\ref{emmh}). 
The dynamical variables we consider are the autocorrelation functions of 
the total-spin and N\'eel-vector operators, 
\begin{equation}
{\cal S}_{\alpha \alpha} (t) = \langle S_{\alpha} (t) S_{\alpha} (0) \rangle
\label{etscf}
\end{equation}
and 
\begin{equation}
{\cal N}_{\alpha \alpha} (t) = \langle n_{\alpha} (t) n_{\alpha} (0) \rangle, 
\label{envcf}
\end{equation}
from which one may seek temporal oscillations characteristic of coherent 
tunnelling dynamics. Working with correlation functions of total spin and 
N\'eel vector is advantageous because it allows one to retain some of 
the spatial symmetries. This simplifies the computation of matrix 
elements of ${\bf S}$ and ${\bf n}$ required in addition to the energy 
spectra for a full understanding of the dynamics. 

The dynamical response of a single spin \cite{rml1} may also be considered 
directly. However, we state that all relevant dynamical properties are encoded 
in the two correlation functions above [Eqs.~(\ref{etscf},\ref{envcf})], and 
show that the single-spin quantities can be deduced from these as follows. 
We denote by $e^{i p}$ the eigenvalue of the one-site translation operator 
on the ring, and presume that all low-energy states are contained in the
sectors $p = 0$ and $p = \pi$, a fact we will verify below. If the 
ground state, $|0 \rangle$, is in the sector $p = 0$, then matrix elements 
$\langle i| S_{\alpha} |0 \rangle$ are finite only for states $|i 
\rangle$ in the sector $p = 0$ and elements $\langle i| n_{\alpha} 
|0 \rangle$ are finite only for states $|i \rangle$ in the sector $p = 
\pi$. Because all states considered are invariant under translations by 
two lattice sites, ${\bf S}$ may be substituted by ${\textstyle 
\frac{1}{2}} N ({\bf s}_1 + {\bf s}_2)$ and ${\bf n}$ by  
${\textstyle \frac{1}{2s}} ({\bf s}_1 - {\bf s}_2)$. It follows that 
single-spin correlation functions are given at low frequencies by 
\begin{equation}
\langle s_{1 \alpha} (t) s_{1 \alpha} (0) \rangle \simeq s^2 {\cal 
N}_{\alpha \alpha} (t) + \frac{1}{N^2} {\cal S}_{\alpha \alpha} (t) ,  
\label{esscf}
\end{equation}
because the cross-correlation functions $\langle n_{\alpha} (t) S_{\alpha} 
(0) \rangle$ vanish due to the opposing symmetries of ${\bf n}$ and ${\bf 
S}$ under one-site translation.

Taking as a guide the semiclassical treatment of Ref.\ \cite{rcl}, 
a tunnelling scenario is applicable if the ground and first excited 
states, $|0 \rangle$ and $|1 \rangle$, are energetically well separated 
from all other states and form a (weakly) tunnel-split doublet with  
splitting $\Delta = E_1 - E_0$. Because the states $|0 \rangle$ and $|1 
\rangle$ have opposite behaviour under translation, as defined above, 
the total-spin matrix element vanishes in this manifold, $\langle 1| {\bf 
S} |0 \rangle = 0$, and ${\cal S}_{zz} (t)$ has no coherent oscillations 
with period characteristic of the tunnelling time $\Delta^{-1}$. By contrast, 
the dynamical properties of the N{\'e}el vector are dominated by tunnelling 
in the lowest manifold, $|\langle 1| n_z | 0 \rangle| \sim 1$, so that 
${\cal N}_{zz} (t)$ should exhibit coherent oscillations with period $2 
\pi/\Delta$. The quantities ${\cal S}$ and ${\cal N}$ are measured through 
the susceptibilities $\chi_{\rm S}$ and $\chi_{\rm N}$, which are directly 
related to Eqs.~(\ref{etscf},\ref{envcf}) by the fluctuation-dissipation 
theorem. $\chi_{\rm S}^{\prime\prime}$ is accessible in ESR or 
alternating-current (AC) susceptibility measurements, but, because 
$\chi_{\rm S}^{\prime\prime} (\omega \sim \Delta) = 0$ shows no response 
at the splitting frequency $\Delta$, does not contain any information on 
tunnelling dynamics \cite{rml1,rml2}, by which is meant processes in the 
low-energy sector. $\chi_{\rm N}^{\prime\prime}$ is the quantity which, 
in accordance with semiclassical theory \cite{rcl,rml1}, should show 
oscillatory behaviour due to coherent N\'eel vector tunnelling. In the 
weak tunnelling regime, $\chi_{\rm N}^{\prime\prime} (\omega \sim \Delta) 
\simeq \pi \delta (\omega - \Delta) \tanh (\beta \Delta / 2)$ has a 
delta-function peak at $\Delta$ \cite{rml1}, but its experimental 
observation is not straightforward. 

A microscopic analysis of Eq.~(\ref{emmh}) involves many energy levels 
coupled by potentially large matrix elements of the operators ${\bf S}$ 
and ${\bf n}$ arising from the spin interactions. Qualitatively, the 
system in a transverse magnetic field (Fig.~1) exhibits two degenerate, 
classical spin configurations (obtained by reflection in the ring plane), 
between which the semiclassical approach predicts a tunnelling scenario. 
However, because of the approximations involved in this description, it 
is not clear that a fully quantum mechanical treatment of the ferric wheel 
systems would confirm the presence of two low-lying levels sufficiently 
well separated from all others, or that ${\mathcal N}_{\alpha \alpha}$ 
would exhibit coherent oscillations dominated by a single frequency for 
any choice of $\alpha$ or of the applied field. We use the exact correlation 
functions, meaning the spectra and matrix elements required in their 
calculation, to resolve this issue. 

Magnetisation curves of Heisenberg rings (Eq.~(\ref{emmh}) with $k_z = 0$) 
were computed some time ago using Lanczos ED \cite{rpb}, while recent
computations of dynamical properties \cite{rw} remain very similar in scope.
The situation becomes more complicated in the presence of a non-zero 
single-ion anisotropy, $k_z \ne 0$, and of a magnetic field applied at an 
arbitrary angle to the ${\hat z}$-axis. First, in this case the eigenvectors 
have a non-trivial dependence on the applied field. Second, when $S_z$ is no 
longer a good quantum number, the loss of the associated symmetries in spin 
space causes a substantial increase in the dimensions of the Hilbert spaces. 
With modern computers it is possible to compute ground-state
properties by the Lanczos method without exploiting spatial symmetries, 
at least in the case of Fe$_6$ \cite{rnwzl}. However, as indicated above, 
dynamical studies require in addtion many excited states and the 
corresponding eigenvectors, although some spatial symmetries remain which 
simplify this task. The Lanczos method can also be used for this 
purpose \cite{rcw}, but becomes substantially more involved. 

The complete Hilbert spaces for Fe$_6$ and Fe$_8$ rings contain $6^6 = 
46656$ and $6^8 = 1679616$ states, respectively.  For the general case 
of Eq.~(\ref{emmh}), there are no symmetries in spin space \cite{footnote}, 
but the spatial symmetries of one-site translation and reflection at a 
given site are present. The ground state in a magnetic field is always 
located either in the sector with $p=0$ or with $p=\pi$, and has positive 
parity with respect to reflection at the selected site. These two sectors 
also contain the lowest excitations. The dimensions of these sectors are 
4291 (107331) for $p=0$ and 4145 (106680) for $p=\pi$ in the case of 
Fe$_6$ (Fe$_8$). Since the N\'eel vector connects these two subspaces, 
we work in the sum of the two spaces, which has dimension 8436 (214011), 
for the computation of matrix elements and correlation functions. 

There are several viable methods for the computation of a large number of 
extremal eigenvalues and -vectors. We have used a combination of simultaneous 
vector iteration of a large number of vectors with explicit diagonalisation 
of the Hamiltonian $H$ in the subspace spanned by these vectors during the 
iteration \cite{rhurl}. For Fe$_6$ we have computed the lowest 350 
eigenvectors and eigenvalues in the symmetry subspace described above. 
The error caused by this truncation may be estimated in two ways. One 
is to examine how the results change on varying the number of states 
retained, and the other is an exact evaluation at $t = 0$ of the 
correlation functions (\ref{etscf}) and (\ref{envcf}) for comparison 
with the results obtained using the truncated spectral representation. 
On the basis of both methods we estimate that the truncation leads to 
errors in the temporal correlation functions for Fe$_6$ on the order of 
$10^{-6}$ of the peak values, meaning that the truncation error is 
undetectable in any of the figures to be shown below for Fe$_6$, while 
numerical errors for the individual eigenstates are considerably smaller 
still. For Fe$_8$ we have retained the lowest 50 eigenvectors in the 
combined subspace, and have an estimated truncation error on the order 
of $10^{-4}$ of the peak values in the temporal correlation functions.
As noted above, all of our results are obtained for zero temperature 
using the minimal Hamiltonian (\ref{emmh}), and thus include no sources 
of decoherence. 

As in Ref.~\cite{rnwzl} we will focus on magnetic fields applied 
in the plane of the ring (Fig.~1), the geometry which retains the 
highest tunnel barrier, with maximal localisation of the N\'eel vector, 
at given field. We have also considered field angles out of the plane of 
the ring, and confirmed that they yield qualitatively similar results at 
strong fields. This is to be expected because the physical situation 
remains one in which the spins are largely confined to a planar motion 
between two potential minima, and also because in our numerical approach 
the $z$-inversion symmetry of the transverse-field case is not essential.
Here we calculate dynamical quantities [Eqs.~(\ref{etscf},\ref{envcf})] 
exhibiting oscillations in the time domain, and discuss the information 
they contain concerning mesoscopic quantum coherent phenomena. 

\begin{figure}[t!]
\centerline{\psfig{figure=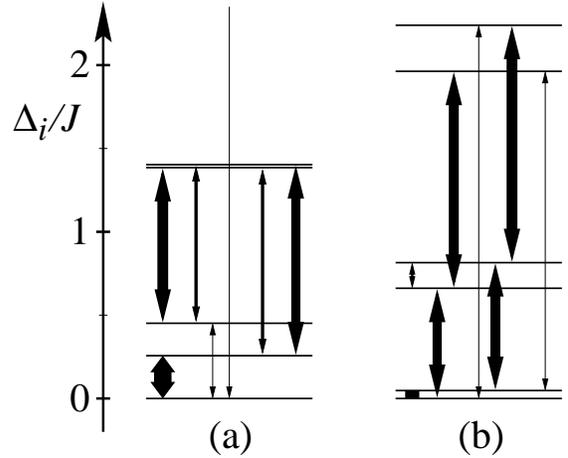,height=6.0cm,angle=0}}
\bigskip
\caption{Lowest energy level spacings $\Delta_i = E_i - E_0$ for an Na:Fe$_6$ 
ring described by the minimal Hamiltonian (\protect{\ref{emmh}}), for applied 
fields $g \mu_B B_x = 3.1 J$ (a) and $g \mu_B B_x = 3.5 J$ (b). The thickness 
of the vertical lines represents the magnitude of the matrix elements of 
$n_z$ connecting each level pair. }
\end{figure}

\section{Results}

\label{secIII}

\subsection{\mbox{\boldmath ${\rm Na:Fe}_6$}}

We begin with a discussion of the energy spectrum for the physical system 
Na:Fe$_6$, for which $k_z/J = 0.0136$ in Eq.~(\ref{emmh}) \cite{rnwzl}. 
As the applied field is increased, level crossings occur at critical fields 
$B_{cn}$ between ground states with increasing total spin and alternating 
quantum numbers $p=0$ or $p=\pi$. This leads to a magnetisation curve with 
an almost regular staircase of plateaux \cite{rnwzl}. Guided by the 
semiclassical prescription that quantum tunnelling is best defined in 
intermediate fields \cite{rcl}, we consider magnetic fields beyond the 
lowest magnetisation plateau ($B > B_{c1}$). 

Fig.~2 shows the lowest energy levels for two fields chosen near the 
centre of a magnetisation plateau (a) and very close to a level crossing 
(b) (see Figs.~5 and 6 below). Near the level crossing, there are indeed 
two nearly degenerate levels lying well below any of the others, but we 
stress that this alone is not sufficient to guarantee single-frequency 
oscillations corresponding to the energy difference $\Delta$ in any 
observable, and thus to justify a two-level tunnelling scenario. This 
point is represented schematically in Fig.~2 by the lines connecting 
the levels, the thickness of which corresponds to the magnitude of the 
matrix element $| \langle i | n_z | j \rangle |$, where $i$ and $j$ 
denote the energy levels. For fields corresponding to the centre of a 
plateau, one observes by contrast that there is no clear two-level 
manifold, but that for certain operators, such as $n_z$ as shown, the 
matrix element between the lowest pair of levels is dominant. This 
situation, $|\langle 1 | n_z | 0 \rangle | \sim 1$, can be taken to 
express the requirement for a two-level description to be adequate 
for mesoscopic tunnelling of the N{\'e}el vector.

\subsubsection{Time Domain}

\label{secIIIA1}

\begin{figure}[t!]
\centerline{\psfig{figure=zf3.eps,width=7.3cm,angle=0}}
\bigskip
\caption{The N\'eel vector correlation function ${\cal N}_{zz} (t)$ for 
Na:Fe$_6$ in magnetic fields (a) $g \mu_B B_x$ = 3.1$J$ and (b) $g \mu_B 
B_x$ = 3.5$J$. }
\end{figure}

\begin{figure}[t!]
\centerline{\psfig{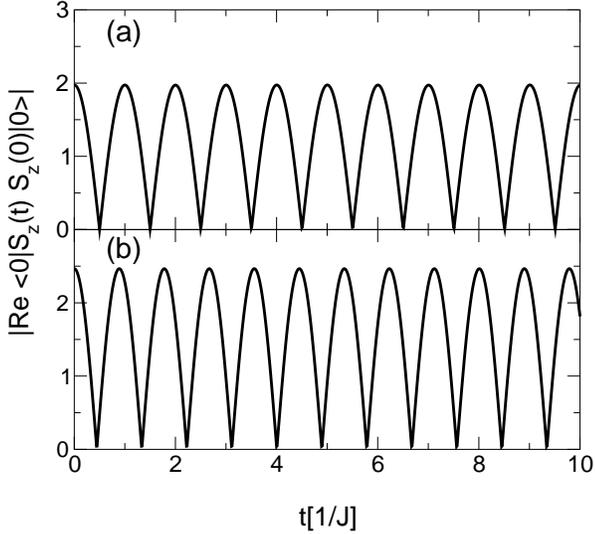}}
\bigskip
\caption{The total-spin correlation function ${\cal S}_{zz} (t)$ for Na:Fe$_6$ 
in magnetic fields (a) $g \mu_B B_x$ = 3.1$J$ and (b) $g \mu_B B_x$ = 3.5$J$.}
\end{figure}

There are in principle two ways of testing the coherent low-energy dynamics 
of a two-level system at low temperatures. The first would be to prepare 
the system in a non-eigenstate of the Hamiltonian, $|\psi \rangle = (|0 
\rangle + |1 \rangle)/\sqrt{2}$, and then to observe coherent oscillations 
of the quantity of interest (here $n_z$) in the time domain. The second is 
to measure ground-state correlation functions such as $ \langle 0|n_z(t) 
\, n_z (0) |0 \rangle$. For an idealised tunnelling scenario in which $n_z$ 
connects only $|0\rangle$ and $|1\rangle$, these two quantities contain 
the same information because in this case 
\begin{equation}
|{\rm Re} \, \langle 0| n_z(t) \, n_z (0) |0 \rangle| \approx | 
\langle \psi| n_z (t) \, n_z (0) | \psi \rangle|.
\label{ecfe}
\end{equation}
In the tunnelling limit, where $|\langle 0| n_z |1 \rangle| \sim 1$, one  
obtains $\langle 0| n_z (t)\, n_z (0) |0 \rangle \simeq |\langle 0| n_z |1 
\rangle|^2 e^{i \Delta t}$. The correlation functions obtained from ED indeed 
show coherent oscillations with periods $2 \pi / \Delta = 24.5/J$ near the 
plateau centre [Fig. 3(a)] and $2 \pi / \Delta = 132.0/J$ near the level 
crossing, where in addition a strong component of a higher harmonic is 
clearly evident [Fig.~3(b)]. The solid and dashed curves in Fig.~3, 
representing respectively the left- and right-hand sides of the two-level 
approximation (\ref{ecfe}), do not coincide because of additional components 
present in the correlation function of $|\psi\rangle$ [right-hand side of 
(\ref{ecfe})]. These indicate that $|1 \rangle$ has significant matrix 
elements of $n_z$ with states other than $|0 \rangle$ (Fig.~2), as a result 
of which the mapping of Na:Fe$_6$ onto a two-level system is marginal (below). 

\begin{figure}[t!]
\centerline{\psfig{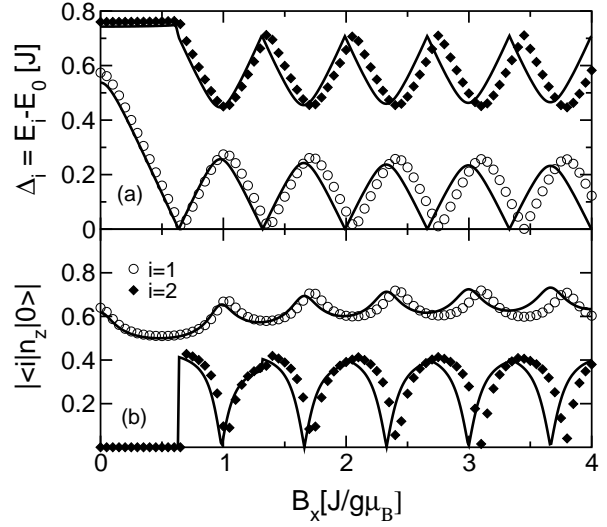}}
\medskip
\caption{Evolution with magnetic field of (a) the energy-level splittings
$\Delta_i$ and (b) the matrix elements $\langle i| n_z |0 \rangle$ for the
low-energy sector ($i = 1,2$) in Na:Fe$_6$. ED results (symbols) are compared
with the RR approximation (lines). }
\end{figure}

In stark contrast to ${\cal N}_{zz}$, the total-spin correlation function 
${\cal S}_{zz}$ (Fig.~4) shows oscillations only at the much higher frequency 
$h_x / 2 \pi$, as expected from the symmetry considerations presented above. 
While these field-driven oscillations are coherent, they do not correspond 
to a tunnelling process, where the levels involved lie below the height of the 
anisotropy barrier, as is the case in the lowest manifold. 

\subsubsection{Spectra and Matrix Elements}

The time-dependent correlation functions shown above may be understood 
directly from the matrix elements between the lowest-lying energy levels 
for the components of the total-spin and N\'eel-vector operators. The 
symbols in Fig.~5 show the numerical results for energy-level splittings 
and matrix elements of $n_z$ in the low-energy manifold of Na:Fe$_6$. The 
energy separations $\Delta_i$ [Fig.~5(a)] confirm that an appreciable 
separation remains between the lowest pair of states and the next higher 
level for all fields. That this criterion alone is not sufficient to assess 
the quality of a two-level description is shown by the matrix elements in 
Fig.~5(b). While $\langle 1|n_z |0 \rangle$ is indeed large for all fields, 
the matrix element $\langle 2|n_z |0 \rangle$ is also significant at fields 
close to the level crossings. In fact at the plateau centers $\langle 
2|n_z| 0 \rangle$ vanishes identically, and in this field regime $\langle 
1|n_z| 0 \rangle$ is considerably larger than $\langle i|n_z |0 \rangle$ 
for all $i \ge 2$. Our ED calculations confirm that the semiclassical 
picture of a tunnelling, or coherent oscillation at sub-barrier energies 
of the N\'eel vector between directions $+{\hat e}_z$ to $-{\hat e}_z$, is 
indeed appropriate here. By contrast, at the level-crossing fields more of 
the higher matrix elements of $n_z$ are appreciable [Fig.~2(b)], and in 
particular $\langle 2|n_z| 0 \rangle$ has 66.0\% of the magnitude of 
$\langle 1|n_z| 0 \rangle$. In the semiclassical description this 
corresponds to the N\'eel vector being rather less well localised along 
$\pm {\hat e}_z$ than for fields at the plateau centers. The effects of 
the higher matrix elements with $|0 \rangle$ [Fig.~5(b)] are clear in the 
difference between the solid lines in Figs.~3(a) and (b), while those of 
the elements with $|1 \rangle$ are visible in the differences between 
solid and dashed lines in Fig.~3. 

\begin{figure}[t!]
\centerline{\psfig{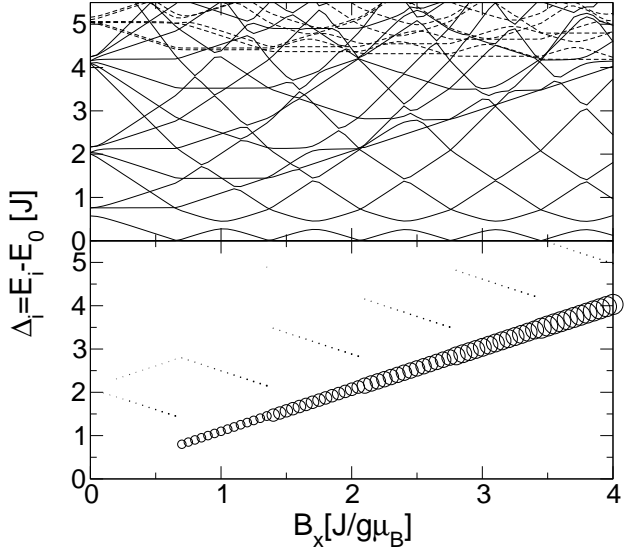}}
\bigskip
\caption{(a) Evolution with magnetic field of energy spectra for Na:Fe$_6$ 
up to $\Delta_i = 5.5J$. The solid lines represent ED data in the momentum 
sectors $p = 0$ and $p = \pi$ (in both cases, only states with positive site 
parity appear in the energy range of the figure), while the dashed lines 
correspond to momenta $p = \pi/3$ and $p = 2 \pi/3$ which are absent in 
the RR model. Note the almost linearly Zeeman-split state at $\Delta_i 
\approx h_x$. (b) Matrix element $|\langle i| S_z |0 \rangle|$ as a function
of magnetic field, represented by radius of open circles. The dominant 
matrix element corresponds to the level with $\Delta_i \approx h_x$ for 
$h_x \ge 0.7 J$, while for the same field range the next-largest elements 
correspond to still higher excited levels. }
\end{figure}

Fig.~6(a) shows the spectrum of Na:Fe$_6$ as a function of field, expanded 
to splittings $\Delta_i = 5.5J$, which illustrates both the predominance of 
the anticrossing between the second and third levels and the presence of 
a linearly evolving Zeeman-split level at $\Delta_i \approx h_x$. The 
total-spin matrix elements in Fig.~6(b) show the two primary features expected 
on symmetry grounds: the matrix element in the low-energy sector vanishes, 
$\langle 1 | S_z | 0 \rangle = 0$, and the dominant matrix element appears 
at the Zeeman splitting. This situation remains very close to the $k_z = 0$ 
limit, where the field-driven Zeeman transition is the only process with a 
non-vanishing matrix element, $|\langle i | S_z | 0 \rangle | = \sqrt{S(h_x) 
/ 2}$, in which $S(h_x)$ denotes the spin of the ground state at field $h_x$. 

To clarify the issue of the importance of choosing $n_z$ and $S_z$, we 
comment that we have also studied correlation functions of the other 
components of ${\bf n}$ and ${\bf S}$. These quantities confirm that 
oscillations generically similar to those observed in ${\cal N}_{zz}$ 
and ${\cal S}_{zz}$ remain. However, the correlation functions of the 
vectors tend to show less clearly defined oscillations as compared to 
those of the $z$-components, because the transverse components couple 
rather more strongly to further levels.

\subsubsection{Rigid-Rotor Model}

Further insight may be gained into the nature of our exact results by 
comparison with those from semiclassical approaches \cite{rldg,rcl,rnwzl},
which specify the conditions for the two-level system to provide an 
appropriate description of the low-energy spectrum of Eq.~(\ref{emmh}). 
Under the assumptions that the spins in the ferric wheel have alternating 
(N\'eel) alignment and that magnon excitations may be neglected, $H$ can 
be mapped to the Hamiltonian of a rigid rotor (RR) \cite{rcl},
\begin{equation}
H_{\rm RR} = \frac{2J}{N} {\bf L}^2 + {\bf h}\cdot {\bf L} - N k_z 
s^2 n_z^2,
\label{ehrr}
\end {equation} 
where ${\bf n}$ and ${\bf L}$ are respectively the position and angular 
momentum of a particle confined to the unit sphere. The operator for total 
spin is represented by the angular momentum of the particle, ${\bf S} = 
{\bf L}$, the eigenstates of which, $|l,m\rangle$, are spherical harmonics. 
The term $- N k_z s^2 n_z^2$ accounts for the anisotropy potential which 
renders energetically favourable those spin configurations for which the 
N\'eel vector ${\bf n}$ is aligned with $\pm {\hat e}_z$. For small 
$k_z$, the eigenstates of Eq.~(\ref{ehrr}) have almost 
uniform probability distribution of ${\bf n}$ in the plane perpendicular 
to the magnetic field, which corresponds to the kinetic limit of $H_{\rm 
RR}$. The opposite limit of large $k_z$ is specified by the condition that 
the tunnel action ${\mathcal S}_0 = N s \sqrt{2 k_z/J}$ be very much 
greater than 1 \cite{rcl}, and it is here that a two-level description 
of quantum tunnelling of the staggered magnetisation is appropriate. 

The condition ${\mathcal S}_0 > 1$ may be taken to mark the onset of a 
spin quantum tunnelling regime, in which there is only one pair of 
tunnel-split states in the low-energy sector. In this respect a two-level 
approximation is marginal for the real materials, although should be rather 
better defined for Cs:Fe$_8$ (${\mathcal S}_0 = 3.8$) and Fe$_{10}$ 
(${\mathcal S}_0 = 3.3$) than for Na:Fe$_6$ (${\mathcal S}_0 = 2.5$). 
In this regime of intermediate $k_z/J$ the RR framework remains 
applicable, but the anisotropy energy $- N k_z s^2 n_z^2$, which may not 
be treated perturbatively, gives rise to significant mixing of states of 
differing angular momentum $l$. A quantitative comparison of RR
results with ED then requires exact diagonalisation of Eq.~(\ref{ehrr}). 
This is most easily performed in the basis $|l,m\rangle$ in which
${\bf L}$ is diagonal, and the matrix elements $\langle l^\prime, m^\prime
| n_z^2 |l,m \rangle$ are evaluated in spherical coordinates. For moderate 
fields, the unphysical states of large angular momenta ($l > 15$) may be 
neglected in the RR approach, and the dimension of the Hilbert space is 
then strongly reduced from that of the full Hamiltonian ($256$ compared to 
$46656$ for Fe$_6$).

Comparisons between ED of the RR model (\ref{ehrr}) and the exact numerical 
results are shown in Fig.~5. There is rather good general agreement in the 
low-energy sector, especially in magnitudes of $\Delta_i$ and $|\langle i| 
n_z |0 \rangle|$, but also a drift in field of the predicted magnetisation 
step positions from the exact result \cite{rsl}. This is thought to be 
largely a consequence of neglecting magnon excitations, by which is meant 
spin misalignments within each of the sublattices, and can be removed by a 
uniform rescaling (with a factor of 1.036 for the parameters of Na:Fe$_6$). 
The RR model is not expected to perform as well at higher energies, a 
statement which can be quantified by inspection of the exact spectra shown 
in Fig.~6(a), where energy levels corresponding to the neglected momentum 
sectors $p = \pi/3$ and $2 \pi/3$ appear at $\Delta_i > 4 J$. However, the 
RR prediction of the total-spin matrix elements shown in Fig.~6(b) remains 
quite accurate for the energies shown, including the qualitative result 
that no matrix elements of the total-spin components connect $|0 \rangle$ 
and $|1 \rangle$ and the quantitative result for the ground-state 
spin $S(h_x) \sim \lfloor N h_x / 4 J \rfloor$ appearing in the only  
large matrix element $|\langle i | S_z | 0 \rangle | \approx \sqrt{S(h_x) 
/ 2}$ at $\Delta_i \approx h_x$. We may summarise by remarking that, for 
${\mathcal S}_0 > 1$, the two-level paradigm \cite{rcl,rml1} delivers a 
simple conceptual picture of the low-energy spin dynamics in terms of 
coherent tunnelling of the staggered magnetisation, and that in addition 
the semiclassical treatment based on ED of the RR approximation provides 
semi-quantitative accuracy for the physical parameters of the real materials. 

\subsection{\mbox{\boldmath ${\rm Cs:Fe}_8$}}

We turn briefly to the 8-membered ferric wheel Cs:Fe$_8$. The effective 
uniaxial anisotropy which may be extracted from the magnetisation data 
for this material \cite{rwskmbsaga} is considerably stronger than in the 
case of Na:Fe$_6$ presented above. Numerically, dynamical simulations 
remain possible for this system, despite the much larger Hilbert space. 
Fig.~7 shows the energy spectrum for a Cs:Fe$_8$ system, again described 
by Eq.~(\ref{emmh}), with the anisotropy ratio deduced \cite{rnwzl} 
from the angle-dependence of the first critical field \cite{rwskmbsaga} to be 
$k_z/J = 0.0185$. The situation remains qualitatively similar to Na:Fe$_6$, 
but has visible differences as a result of the larger values of $N$ and 
$k_z/J$. The stronger coupling between eigenstates $|l,m \rangle$ results 
in a stronger anticrossing of levels $|1 \rangle$ and $|2 \rangle$ at the 
plateau centers and a smaller maximal $\Delta$. This enhanced separation 
of the two-state manifold at lowest energies, combined with an increased 
tunnel barrier, makes the semiclassical tunnelling description more 
appropriate, as expected from the larger value of ${\cal S}_0$. Although 
in Fig.~7 we show only the results of full diagonalisation, we have verifed 
that, as in the case of Na:Fe$_6$, the RR model again provides qualitatively 
similar results, albeit with a larger drift in the level-crossing fields. 

\begin{figure}[t!]
\centerline{\psfig{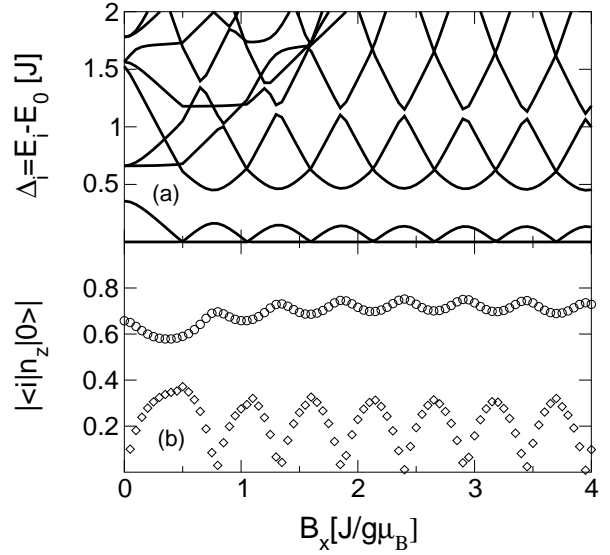}}
\bigskip
\caption{Evolution with magnetic field of (a) energy-level splittings
$\Delta_i$ and (b) matrix elements $\langle i| n_z |0 \rangle$ for the
low-energy sector in Cs:Fe$_8$. In (b), the circles correspond to $i = 1$
and diamonds to $i \ge 2$.}
\end{figure}

\begin{figure}[t!]
\centerline{\psfig{figure=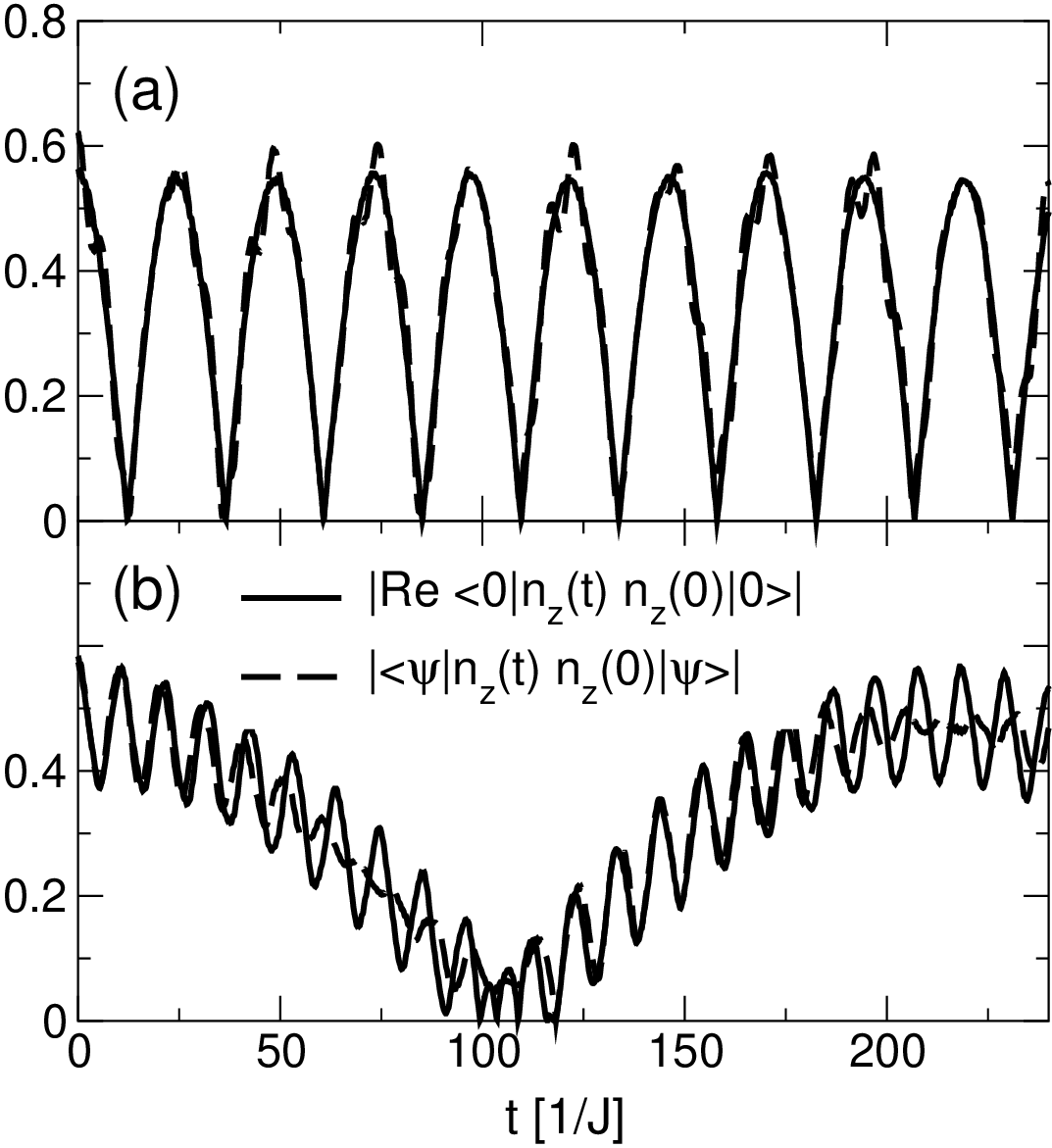,width=7.0cm,angle=0}}
\bigskip
\caption{The N\'eel vector correlation function ${\cal N}_{zz} (t)$ for 
Cs:Fe$_8$ in magnetic fields (a) $g \mu_B B_x$ = 3.4$J$ and (b) $g \mu_B 
B_x$ = 3.72$J$. }
\end{figure}

Fig.~8 shows the N\'eel vector correlation function for the system in two 
magnetic fields chosen close to a plateau centre and to a plateau edge. We 
observe rather clean, single-component temporal oscillations near the centre 
of the magnetisation plateau [Fig.~8(a)], where the admixture of higher 
frequencies is indeed weak. Near the level crossing [Fig.~8(b)] the 
second-largest matrix element, $\langle 2|n_z|0 \rangle$, has 43.4\% of 
the magnitude of $\langle 1|n_z|0 \rangle$, demonstrating again the reduced 
validity of the two-level description at these fields, compared to the 
plateau centers. However, the large separation in frequency scales results 
in clearly visible contributions from both components, and thus to 
pronounced low-frequency oscillations in ${\cal N}_{zz}$ corresponding to 
coherent N\'eel vector tunnelling in Fig.~8(b).

\section{Discussion}

\label{secIV}

\subsection{Nature of mesoscopic quantum dynamics}  

With the aid of the two-level analogue we obtain a conceptual picture of the 
microscopic nature of quantum coherence. The main result of our ED study
is that the correlation functions of the total spin and the N\'eel
vector, ${\mathcal S}_{\alpha \alpha}(t)$ and ${\mathcal N}_{\alpha 
\alpha}(t)$ respectively, show qualitatively different behaviour. Although 
both exhibit almost single- (or two-)frequency oscillations for the parameters 
illustrated here, the frequencies of these oscillations are very different. 
While ${\mathcal S}_{zz}(t)$ oscillates at a frequency determined by the 
(strong) magnetic field, $\omega = h_x$, ${\mathcal N}_{zz}(t)$ exhibits 
oscillations with a much smaller frequency, $\omega = \Delta$. The amplitude 
of the N\'eel-vector correlation function, ${\mathcal N}_{zz}(t) \lesssim 1$, 
also allows one to conclude that the oscillations indeed correspond to a 
true quantum tunnelling. Hence, although most features of the thermodynamic 
properties and the ESR spectra of ferric wheels can be understood from a 
phenomenological total-spin Hamiltonian \cite{rtdpfgl,rcja}, their most 
interesting dynamical feature, quantum coherent spin tunnelling, is not 
contained in such a description.

The clearest example of spin tunnelling would be obtained by preparing a 
two-level system in the state $|\psi \rangle$ introduced in
Sec.~\ref{secIIIA1}. 
For a ferric wheel this state would correspond to a spin configuration of 
the type represented in Fig.~1, with the N\'eel vector oriented in the 
direction $+{\hat e}_z$. $|\psi \rangle$ is a superposition of energy 
eigenstates whose evolution proceeds according to the phase factors 
$e^{- i E_i t}$ of the energy levels, without disturbance by external 
processes (decoherence), such that after every odd number of half cycles 
the degenerate state $|{\bar \psi} \rangle = (|0 \rangle - |1 \rangle) / 
{\sqrt 2}$, corresponding to the N\'eel vector oriented along $-{\hat 
e}_z$, is achieved. 

In the microscopic analysis of the ferric wheel, we have demonstrated two 
important features in the realisation of such an idealised tunnelling scenario. 
First, it is indeed possible to reduce the complex physical system to an 
effective two-state model, by application of a strong field in the ring 
plane. In this case, the dominant matrix element of the selected operator 
$n_z$ is that connecting the two levels in the low-energy manifold, such 
that ${\mathcal N}_{zz}$ exhibits single-frequency oscillations at the 
splitting frequency $\Delta$, which corresponds to tunnelling dynamics 
of the N\'eel vector. We stress in this connection that single-frequency 
oscillations depend not only on the energetic separation of the levels 
involved (from each other or from all others in the spectrum), but 
essentially on the matrix elements. While other single-frequency 
oscillations may be found in a dissipationless quantum system, for the 
ferric wheel only that in the low-energy manifold corresponds to a tunnelling
process. The second key feature is that in this situation the difficult 
experimental step of establishing the non-eigenstate $|\psi \rangle$ is 
not essential, given a suitable measurement of the equilibrium correlation 
function ${\mathcal N}_{zz}(t)$. We discuss in Sec.~\ref{secIVC} below the 
possibilities for executing this nontrivial task. 

Finally, we comment that the ferric wheels are intermediate in size 
between a minimal quantum system such as two coupled $s = 1$ spins with 
anisotropy and the classical limit of a large total (staggered) spin 
$S \gg 1$ with at least two degenerate minima in an energy 
continuum. Comparison of our exact numerical results with analytical 
expressions for small systems, and with the RR model, provides a 
prescription for understanding the evolution of coherent phenomena with 
system size, and the accompanying evolution of the appropriate description 
from microscopic to semiclassical. 

\subsection{Decoherence}

As already discussed in Sec.~\ref{secI}, we have not included decoherence
within our numerical calculations. On the technical level this can be effected 
by the inclusion of a generic bath coupled to the quantum system \cite{rsmd}. 
However, with the aim of establishing the nature of coherent oscillations 
in the mesoscopic ferric wheel systems, we have restricted the present 
discussion to the pragmatic level on which decoherence times $\Gamma^{-1}$ 
significantly longer than tunnelling times $\Delta^{-1}$ are a prerequisite 
for any observation of coherent phenomena \cite{rgb}. 

In the ferric wheel systems the tunnel splitting frequencies $\Delta$ are 
very large: at the centers of the magnetisation plateaux, where we have shown 
that the clearest single-frequency oscillations should be present, the 
tunnel frequencies are $\Delta = 235$GHz (11.4K) in Na:Fe$_6$, 62GHz (3.0K) 
in Cs:Fe$_8$ and 45GHz (2.2K) in Fe$_{10}$. At low temperatures in a 
sufficiently pure crystal, decoherence will be due primarily to additional
interactions not included in the minimal Hamiltonian of Eq.~(\ref{emmh}). 
The intrinsic decoherence rate, $\Gamma$, for ferric wheels is most likely 
to be controlled by interring dipolar interactions of electron spins 
($10$-$50$mK), and possibly by interring superexchange processes, whose 
contributions are very difficult to estimate but may exceed 0.1K \cite{rml2}. 
Nuclear dipolar interactions with $^1$H nuclei ($0.1$mK) and hyperfine 
interactions with $^{57}$Fe nuclei ($1$mK) are sufficiently weak \cite{rml1} 
that they may be neglected in these systems. An approximate upper 
limit for $\Gamma$ in Na:Fe$_6$ may be taken from the broadening of the 
lowest magnetisation step at 40mK \cite{rceau}, which was measured as 0.4T 
(0.54K), although this is considerably smaller than the broadenings 
observed at the magnetisation steps by NMR relaxation rate measurements 
in Li:Fe$_6$ \cite{rleau}. In any event, even with these worst-case 
decoherence estimates, the ferric wheels Fe$_{10}$, Cs:Fe$_8$ and 
Na:Fe$_6$ remain by the $\Delta \gg \Gamma$ criterion very promising 
candidates for observation of coherent spin tunnelling. 

\subsection{Detection of coherent tunnelling} 

\label{secIVC}

Using the above picture of mesoscopic quantum coherence at the microscopic 
level, one would like to consider those experiments or techniques which 
may be applied in order to detect coherent quantum tunnelling in the ferric 
wheels. Unfortunately, our prescription is at first sight rather exacting, 
as it requires observing time-dependent correlation functions of the N\'eel 
vector. An observation would then depend on a probe coupling to the 
staggered moment of the AF rings. The closest available options for 
dynamical investigations are the local spin-raising and -lowering 
operations ($\Delta s_i^z = \pm 1$) in NMR and INS studies. We have argued 
that single-spin dynamics in the ring indeed reflects the N\'eel vector 
response, and thus conclude that these techniques are in principle capable 
of revealing the existence of coherent tunnelling processes at frequency 
$\Delta$. The matrix elements we have computed could also be used to fit 
transition intensities higher in the INS spectrum. In contrast to these 
probes, ESR is sensitive only to the total spin ${\bf S}$ of the ferric 
wheels, and thus, as shown above and in Ref.~\cite{rml1}, cannot 
detect mesoscopic N\'eel vector tunnelling. The matrix elements for total 
spin indicate that the ESR response would be dominated by the field-driven 
Zeeman transition. 

However, technical problems arise which would make any observation 
of quantum coherence very difficult for the pure ferric wheels which are 
known. NMR measurements suffer from the 
very weak matrix elements coupling the nuclear spins to the electronic 
system \cite{rml1}, and from the mismatch in frequency scales between the
maximal $\Delta$ and the probe frequency. Neutron scattering studies 
would require a large, deuterated single crystal. Here we suggest only 
that the most appropriate way forward would be to expand the range of 
viable experiments by considering rings with a strongly broken symmetry, 
as may be achieved by doping with nonmagnetic ($s = 0$) impurities 
\cite{rml2}, such as Ga in ferric wheels. These modified ferric wheels retain 
the AF properties required to satisfy the condition $\Delta \gg \Gamma$, and 
in addition possess an excess, or tracer, spin which may be followed with a 
magnetic field. The dynamical properties of these systems are expected to 
retain the quantum tunnelling characteristics of the undoped 
systems \cite{rml2}, with the important additional feature that the total 
spin ${\bf S}$ also reflects the coherent tunnelling of ${\bf n}$, which 
would then become accessible by ESR \cite{rml1}. We will address in a 
forthcoming publication the microscopic and experimental aspects of 
dynamics in modified ferric wheels.

\section{Summary}

\label{secV}

In conclusion, we have presented a numerical and semiclassical analysis of 
low-energy spin dynamics and quantum coherent tunnelling phenomena in the 
molecular magnetic ring systems Na:Fe$_6$ and Cs:Fe$_8$. The energy-level 
spectra and the matrix elements for total-spin and N\'eel-vector operators 
computed at different magnetic field values establish the presence of 
quantum coherent oscillations in their correlation functions. Oscillations 
corresponding to coherent tunnelling of the staggered magnetisation are 
present for systems with the physical parameters of the ferric wheels, 
with the cleanest single-frequency, or two-level, oscillations found in 
the response of the N\'eel vector at applied fields in the centers of 
intermediate magnetisation plateaux. These results show that, despite the 
small system size ($N = 6$ or 8) and small spin quantum number $s = 5/2$, 
the semiclassical picture of N\'eel vector tunnelling \cite{rcl} provides 
a valid picture of the low-energy dynamical properties of ferric wheels. 
Experimental observation of mesoscopic quantum oscillations is possible 
with probes which flip local electronic spins, and would appear to be 
most feasible in studies of broken-symmetry systems obtainable by doping 
of the ring materials. 

\section*{Acknowledgements}

FM and DL are supported by the Swiss National Fund, Molnanomag 
HPRN-CT-1999-00012 and BBW Bern. BN is supported by SFB 484 of the 
Deutsche Forschungs\-gemeinschaft.


\begin{thebibliography}{99}
\bibitem{footnote} For a field in the $x$-$y$ plane, $z$-direction spin 
inversion is a further symmetry, which we have, however, not exploited.
\end{thebibliography}
\end{document}